%% file: main.tex
\documentclass[twocolumn, tighten]{aastex62}
\usepackage{xspace}
\usepackage{graphicx}
\usepackage{amssymb}
\usepackage{amsmath}
\usepackage{hyperref}
\usepackage{footmisc}

\usepackage{natbib} 

\newcommand{\myemail}{zpace@astro.wisc.edu}
\newcommand{\Dn}{D$_n$4000\xspace}
\newcommand{\HdeltaA}{H${\delta}_A$\xspace}

\newcommand{\mplv}{MPL-8\xspace}
\newcommand{\mplvngal}{6779\xspace}
\newcommand{\mplvfull}{MaNGA Product Launch 8\xspace}

\newcommand{\nrungalaxies}{6356\xspace}

\newcommand{\logml}[1]{\ensuremath{\log \Upsilon^*_{#1}}}

\defcitealias{pace_19a_pca}{Paper \textsc{i}}
\defcitealias{chen_pca}{C12}

\accepted{29 July 2019}
\submitjournal{ApJ}

\shorttitle{PCA Stellar Masses: Paper \textsc{ii}}
\shortauthors{Pace et al.}

\begin{document}

\title{Resolved and Integrated Stellar Masses in the SDSS-\textsc{iv}/\textsc{MaNGA} Survey, Paper \textsc{ii}: \\ Applications of PCA-based stellar mass estimates}

\correspondingauthor{Zachary J. Pace}
\email{\myemail}

\author[0000-0003-4843-4185]{Zachary J. {Pace}}
\affil{Department of Astronomy, University of Wisconsin-Madison, 
       475 N Charter St., Madison, WI 53706}

\author[0000-0003-3097-5178]{Christy {Tremonti}}
\affil{Department of Astronomy, University of Wisconsin-Madison, 
       475 N Charter St., Madison, WI 53706}

\author{Yanmei {Chen}}
\affil{Department of Astronomy, Nanjing University, 
       Nanjing 210093, China}

\author{Adam L. {Schaefer}}
\affil{Department of Astronomy, University of Wisconsin-Madison, 
       475 N Charter St., Madison, WI 53706}

\author[0000-0002-3131-4374]{Matthew A. {Bershady}}
\affil{Department of Astronomy, University of Wisconsin-Madison, 
       475 N Charter St., Madison, WI 53706}
\affil{South African Astronomical Observatory, 
       P.O. Box 9, Observatory 7935, Cape Town South Africa}

\author[0000-0003-1809-6920]{Kyle B. {Westfall}}
\affil{UCO/Lick Observatory, University of California, Santa Cruz, 
       1156 High St., Santa Cruz, CA 95064}

\author[0000-0003-0946-6176]{M\'{e}d\'{e}ric Boquien}
\affil{Centro de Astronom\'{i}a, Universidad de Antofagasta, 
       Avenida Angamos 601, Antofagasta 1270300, Chile}

\author[0000-0001-7883-8434]{Kate {Rowlands}}
\affil{Center for Astrophysical Sciences, Department of Physics and Astronomy, 
       Johns Hopkins University \\
       3400 North Charles Street, Baltimore, MD 21218, USA}

\author[0000-0001-8085-5890]{Brett {Andrews}}
\affil{TT PACC, Department of Physics and Astronomy, University of Pittsburgh, 
       Pittsburgh, PA 15260, USA}

\author[0000-0002-8725-1069]{Joel R. {Brownstein}}
\affil{Department of Physics and Astronomy, University of Utah, 
       115 S. 1400 E., Salt Lake City, UT 84112, USA}

\author[0000-0002-7339-3170]{Niv {Drory}}
\affil{McDonald Observatory, The University of Texas at Austin, 
       1 University Station, Austin, TX 78712, USA}

\author[0000-0002-6047-1010]{David {Wake}}
\affil{Department of Physics, University of North Carolina Asheville, 
       One University Heights, Asheville, NC 28804, USA}



\begin{abstract}

A galaxy’s stellar mass is one of its most fundamental properties, but it remains challenging to measure reliably. With the advent of very large optical spectroscopic surveys, efficient methods that can make use of low signal-to-noise spectra are needed. With this in mind, we created a new software package for estimating effective stellar mass-to-light ratios $\Upsilon^*$ that uses principal component analysis (PCA) basis set to optimize the comparison between observed spectra and a large library of stellar population synthesis models. In \citetalias{pace_19a_pca}, we showed that a with a set of six PCA basis vectors we could faithfully represent most optical spectra from the Mapping Nearby Galaxies at APO (MaNGA) survey; and we tested the accuracy of our M/L estimates using synthetic spectra. Here, we explore sources of systematic error in our mass measurements by comparing our new measurements to data  from the literature. We compare our stellar mass surface density estimates to kinematics-derived dynamical mass surface density measurements from the DiskMass Survey and find some tension between the two which could be resolved if the disk scale-heights used in the kinematic analysis were overestimated by a factor of $\sim1.5$. We formulate an aperture-corrected stellar mass catalog for the MaNGA survey, and compare to previous stellar mass estimates based on multi-band optical photometry, finding typical discrepancies of 0.1 dex. Using the spatially resolved MaNGA data, we evaluate the impact of estimating total stellar masses from spatially unresolved spectra, and we explore how the biases that result from unresolved spectra depend upon the galaxy’s dust extinction and star formation rate. Finally, we describe a SDSS Value-Added Catalog which will include both spatially resolved and total (aperture-corrected) stellar masses for MaNGA galaxies.

\end{abstract}



\input{intro.tex}

\input{data.tex}

\input{method.tex}

\input{discussion.tex}

\input{acknowledgements.tex}

\software{Astropy \citep{astropy}, matplotlib \citep{matplotlib}, statsmodels \citep{seabold2010statsmodels}}

\bibliography{main}


\listofchanges

\end{document}

%% file: intro.tex
\section{Introduction}
\label{sec:intro}

A galaxy's total stellar mass is a helpful indicator of its overall evolutionary state: more massive galaxies tend to reside in older, more massive dark matter haloes \citep{gallazzi_05_age_metallicity, gallazzi_06}, and they tend to have exhausted or expelled the majority of their cold gas in previous generations of star-formation star formation \citep{kauffmann_heckman_white_03, balogh_04_cmd, baldry_06_massquenching}. Similarly, more massive galaxies have stellar populations and gas which are relatively metal-enriched \citep{gallazzi_charlot_05, gallazzi_06}. In contrast, low-mass galaxies are still forming stars from their high mass fraction of cold gas \citep{mcgaugh_de-blok_97}, and continue to enrich their interstellar medium (ISM) from a relatively pristine chemical state \citep{tremonti_mz}.

It is possible to roughly quantify a galaxy's total stellar mass by measuring its stellar mass-to-light ratio $\Upsilon^*$ and then multiplying by the galaxy's luminosity. Two main classes of methods have been employed to make this calculation. The DiskMass Survey \citep[DMS, ][]{diskmass_i}, for instance, measured stellar and gas kinematics for 30 face-on disk galaxies in the local universe, and combined them with estimates of typical disk scale-heights to calculate dynamical mass surface density in several radial bins for each galaxy. Such dynamical measurements are limited, though, by potential systematics in the adopted disk scale-heights and assumptions about the stellar velocity dispersions \citep{aniyan_freeman_16, aniyan_freeman_18}.

An alternative is to compare the light emitted by stars in a galaxy to stellar population synthesis models \citep{tinsley_72, tinsley_73}. This involves wedding theory of stellar evolution to a stellar initial mass function (IMF), a star-formation history, and either theoretical model stellar atmospheres or empirical libraries of stellar photometry or spectroscopy. In general, one can either attempt to match an observed spectral-energy distribution (SED) directly to a star-formation histories (SFH), or use a library of SFHs to quantify, for example, a relationship between an optical color and a stellar mass-to-light ratio \citet{bell_dejong_01, bell_03}. Related approaches for optical spectra \citep{kauffmann_heckman_white_03} rely on measuring stellar spectral indices, such as the 4000-\mbox{\AA} break (\Dn) or the equivalent-width of the H-$\delta$ absorption feature (\HdeltaA). The process of reconstructing the exact SFH that produces an optical spectrum, though, is fraught with degeneracies: broadly speaking, mean stellar age, stellar metallicity, and attenuation due to dust are extremely covariate. Thus, recent efforts have concentrated on reducing the intractability of this problem.

In \citet[][hereafter \citetalias{pace_19a_pca}]{pace_19a_pca}, we applied a relatively innovative approach to inferring a stellar mass-to-light ratio from an optical spectrum, which used a library of $\sim$ 40000 SFHs and the associated optical spectra to construct a more computationally-friendly fitting framework, following the method of \citet[][, hereafter \citetalias{chen_pca}]{chen_pca}. Using principal component analysis (PCA), we computed a set of six basis vectors which were able to faithfully represent most optical spectra from the MaNGA survey\footnote{These six ``eigenspectra" do not themselves directly represent physical quantities; rather, when combined, they can serve to emulate spectra of stellar populations, and so are taken to encode more abstractly quantities of interest (such as stellar mass-to-light ratio, stellar metallicity, or dust attenuation).}. By projecting each observed spectrum down on the basis set of the principal component vectors, and adopting covariate observational uncertainties, we better account for theoretical degeneracies in stellar population synthesis, and obtain estimates of stellar mass-to-light ratio usable at a variety of signal-to-noise ratios, metallicities, and foreground dust attenuations.

Galaxies are not single points of light, though. Stellar populations vary on spatial scales comparable to the sizes of individual H\textsc{ii} regions, and undertaking observations that sample only at kiloparsec (or coarser) resolution therefore mix together light from locations in a galaxy that have very different properties and physical conditions. Thus, some information will necessarily be lost. For example, several studies have found that measurements based on integrated color-mass-to-light relations (CMLRs) tend to underestimate the total stellar mass (by 25--40\%, depending on specific star-formation rate) because dusty regions contribute very little to the integrated colors \citep{zibetti_2009,sorba_sawicki_15,martinez-garcia_17}. Explanations for this phenomenon run the gamut from simple variation in intrinsic luminosity of stellar populations across a galaxy's face (the ``outshining hypothesis") to the spatial variation of dust inducing a simultaneous reduction in total luminosity and increase in effective stellar mass-to-light ratio. Furthermore, the same mechanism need not dominate when transforming between different spatial scales (sub-kiloparsec to several-kiloparsec, or kiloparsec to galaxy-wide).

In this work, we build further on the resolved estimates of stellar mass-to-light ratio and resolved estimates of stellar mass obtained in \citetalias{pace_19a_pca}: In Section \ref{sec:data}, we briefly summarize the SDSS-IV/MaNGA survey and the data products upon which our analysis depends. In Section \ref{sec:dms_compare}, we compare PCA-derived estimates of resolved stellar mass surface-density (SMSD) to those of dynamical mass surface density (DMSD) from the DiskMass Survey. In Section \ref{sec:mstar_catalog}, we obtain a catalog of total galaxy stellar mass after testing two aperture-correction methods designed to account for stellar mass outside the spatial sampling area of the MaNGA instrument. In Section \ref{subsec:resolution_effects}, we examine the effects of spatially-coadding all spectral pixels on the total galaxy stellar mass obtained. Finally, in Section \ref{sec:disc}, we briefly summarize our results and outline the future release of a SDSS value-added catalog (VAC) of total galaxy stellar-masses.

%% file: data.tex
\section{Data}
\label{sec:data}

This work further analyzes principal component analysis (PCA) fits from \citet{pace_19a_pca} to integral-field spectroscopic data from the MaNGA survey \citep{bundy15_manga}, part of SDSS-IV \citep{blanton_17_sdss-iv}. MaNGA is an integral-field survey which targets upwards of 10,000 nearby galaxies ($0.01 < z < 0.15$). The NASA-Sloan Atlas \citep[NSA, ][]{blanton_11_nsa} provides the majority of the targets for the MaNGA survey. Two-thirds of targets are drawn from the ``Primary+" sample, which have spatial coverage to at least 1.5 $R_e$; and the remaining one-third from the ``Secondary" sample, which have spatial coverage to 2.5 $R_e$. In order to obtain an approximately-flat distribution in galaxy $\log M^*$ \citep{manga_sample_wake_17}, MaNGA targets are selected uniformly in $i$-band absolute magnitude \citep{fukugita_96_sdss_photo, doi_2010_sdssresponse}. Within a particular redshift range, the MaNGA sample is also selected to be volume-limited. Absolute magnitudes, tabulated in the \texttt{DRPALL} catalog file, have been calculated using K-corrections computed with the \texttt{kcorrect v4\_2} software package \citep{blanton_roweis_07}, which assumed a \citet{chabrier03} stellar initial mass function and \citet{BC03} SSPs.

MaNGA observations employ the BOSS spectrograph \citep{smee_boss_instrument, sdss_boss_dawson_13}, an instrument on the SDSS 2.5-meter telescope at Apache Point Observatory \citep{gunn_sdss_telescope}. The spectrograph covers the optical--near-IR wavelength range (3600 to 10300 $\mbox{\AA}$) at a spectral resolution $R \sim 2000$. Galaxies are spatially-sampled by coupling the BOSS spectrograph's fiber feed to closely-packed hexabundles of fiber-optic cables, called integral-field units (IFUs), which each have between 19 and 127 fibers \citep{manga_inst}. Each fiber subtends 2" on the sky. Like previous SDSS surveys, hexabundles are affixed to the focal plane using a plugplate, and are exposed simultaneously \citep{sdss_summary}. Sky subtraction relies on 92 single fibers spread across the focal plane. Twelve seven-fiber ``mini-bundles" (six per half-plate) simultaneously observe standard stars, and are used for spectrophotometric flux calibration \citep{manga_spectrophot}.

MaNGA data are provided in both row-stacked spectra (RSS) and datacube (LINCUBE \& LOGCUBE) formats. RSS exposures are rectified into a datacube using a modified Shepard's algorithm, such that the size of the spatial element (spaxel) is 0.5" by 0.5" \citep{manga_drp}. The LOGCUBE products have logarithmic wavelength spacing ($d\log \lambda = 10^{-4}$, $d\ln \lambda \approx 2.3 \times 10^{-4}$)\footnote{In this work, the notation $\log$ denotes a base-10 logarithm, and $\ln$ denotes a base-$e$ logarithm.}. The dithering approach ensures that 99\% of the face of the target object is exposed to within 1.2 \% of the target depth \citep{manga_obs}. Sets of three exposures are accumulated until a threshold signal to noise ratio is achieved \citep{manga_progress_yan_16}. The typical point-spread function of a MaNGA datacube has a FWHM of 2.5" \citep{manga_obs}. The MaNGA Data Analysis Pipeline \citep[DAP, ][]{manga_dap} measures stellar kinematics, emission-line strengths, and stellar spectral indices for individual spaxels. The PCA analysis undertaken in \citetalias[][, the results of which we employ here]{pace_19a_pca} relies on the DAP products to establish a velocity field and deredshift individual spectra into the rest-frame.

This work builds on \citetalias{pace_19a_pca}, which inferred stellar mass-to-light ratio using a PCA-based spectral-fitting and parameter estimation method similar to earlier work from the SDSS-III/BOSS survey \citepalias{chen_pca}. This analysis relied upon a set of composite stellar populations (CSPs), whose optical spectra were used to construct a lower-dimensional spectral-fitting basis set. The spectral-fitting paradigm adopted in \citetalias{pace_19a_pca} takes into better account uncertainties in stellar population synthesis (such as the age-metallicity degeneracy), along with covariate uncertainty accounting for imperfect spectrophotometry. \citetalias{pace_19a_pca} analyzed \nrungalaxies galaxies drawn randomly from \mplvfull (\mplv), a set of \mplvngal observations of galaxies taken between March 2014 and June 2018, and returned resolved estimates \& their uncertainties for quantities of interest such as $i$-band effective stellar mass-to-light ratio, \logml{i}. \citetalias{pace_19a_pca} established the quality of the \logml{i} estimates by vetting them against synthetic observations of held-out test models: PCA-based estimates of \logml{i} were found to have on average very modest ($\lesssim 0.1~{\rm dex}$) systematics over a wide range of optical colors, stellar metallicites, and foreground dust attenuations. The estimates of uncertainty for \logml{i} were also found to be realistic.

%% file: method.tex
\section{Resolved stellar-mass surface densities: comparison to results from dynamics}
\label{sec:dms_compare}

While estimates of stellar mass-to-light ratio based on SPS do have systematics (e.g., relating to the prevalence and strength of starbursts---see \citetalias{pace_19a_pca}), those systematics differ in important ways from those affecting dynamical measurements of stellar-mass. Speaking generally, the measured quantity in dynamical studies is the dynamical mass surface density (DMSD), which is proportional to the square of the ratio between vertical velocity dispersion $\sigma_z$ and disk scale-height $h_z$ \citep{diskmass_i}. Measuring $\sigma_z$ requires either a perfectly face-on galaxy (rare) or a decomposition of the line-of-sight velocity distribution (LOSVD) for a moderately-inclined galaxy. For moderately-inclined disks, LOSVD decomposition generally relies on a rotation curve from cold or warm gas (\textsc{Hi} or H$\alpha$), which arise most often in galaxies with a mixture of young and old stellar populations (the younger being dynamically colder). Furthermore, as \citet{aniyan_freeman_18} notes, $\sigma_z$ and $h_z$ must reflect the same stellar population. \citet{aniyan_freeman_16} concluded that a factor-of-two underestimate of DMSD could arise if the two kinematical populations are conflated into one.

With these considerations in mind, it would be helpful to evaluate the two methods of mass determination in direct contrast. MaNGA, for instance, has observed three galaxies also part of the DiskMass Survey \citep[DMS, ][]{diskmass_i}: DMS measured dynamical mass surface density in several radial bins for UGC3997 (8566-12705), UGC4107 (8567-12701), and UGC4368 (8939-12704). For all of these galaxies, the gas fraction by mass is small, and the dark matter distribution contributes minimally to the vertical velocity structure. In other words, the stellar mass should strongly dominate the dynamical mass. Therefore, SMSD estimated from SPS should be just slightly less than the DMSD estimated from kinematics. For a given galaxy, comparing the two radial profiles can provide some intuition for the extent to which the above systematics are important.

In the DiskMass Survey \citep{diskmass_i}, the dynamical mass surface density $\Sigma^{\rm dyn}$ is determined according to the vertical velocity-dispersion $\sigma_z$, disk scale-length $h_R$, and disk scale-height $h_z$:
\begin{equation}
    \Sigma^{\rm dyn} \propto \frac{\sigma_z^2}{h_z}
    \label{eqn:dyn_smsd}
\end{equation}

\begin{figure*}
    \centering
    \includegraphics[height=0.8\textheight]{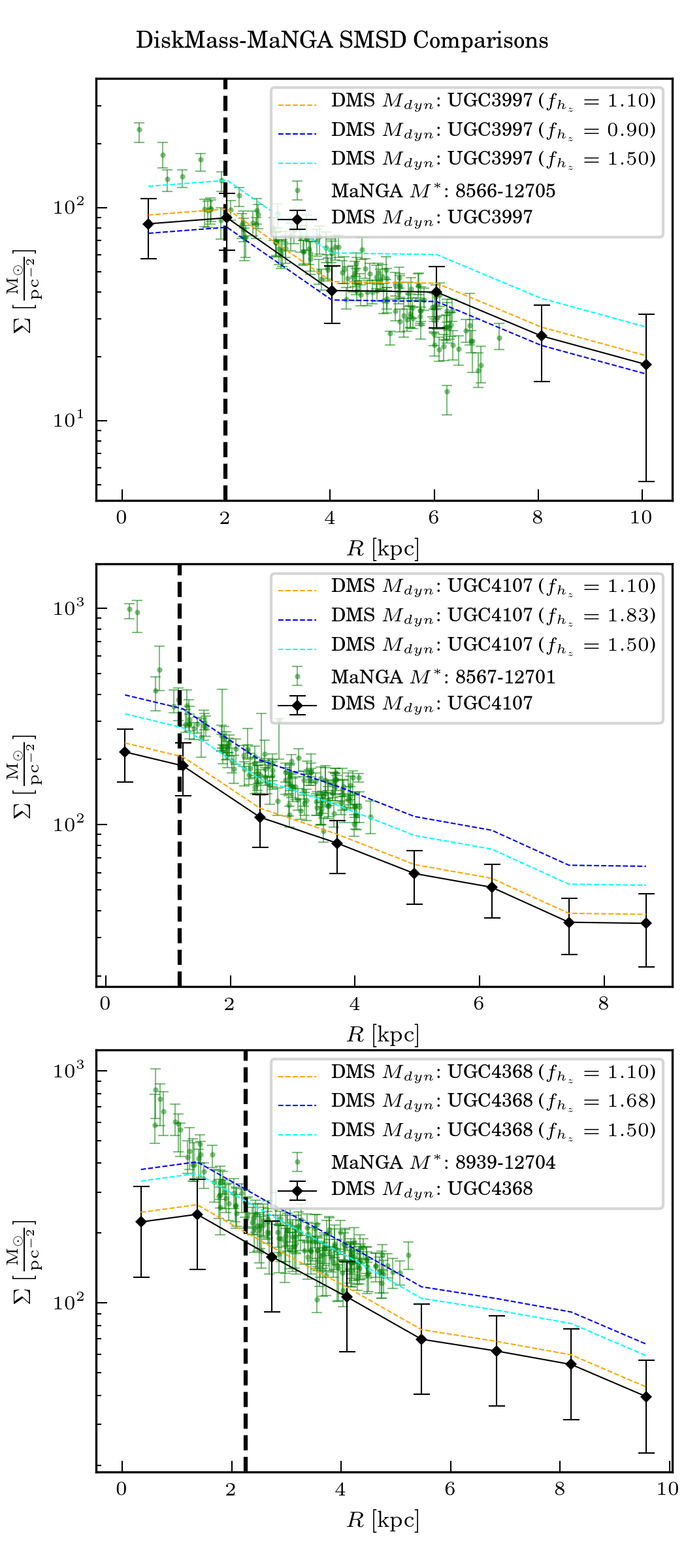}
    \caption{Comparison between PCA-measured stellar mass surface density (green circles) to the dynamical-mass surface-densities (DMSDs) from the DiskMass Survey (black diamonds connected by black, solid line, corrected to our fiducial $h_z$) for three galaxies: 8566-12705 (UGC3997), 8567-12701 (UGC4107), and 8939-12704 (UGC4368). For each galaxy, the radius of the bulge is shown as a vertical black dashed line: bulges differ from disks in their density profile, dynamics, and stellar populations, so the same kinematical relationships do not apply \citep{diskmass_vii}. We also show the values of $\Sigma_{dyn}$ which would result from three different corrections to $h_z$ (i.e., assuming three values of $f_{h_z}$): in orange, the DMSD assuming $f_{h_z} = 1.1$; in cyan, assuming $f_{h_z} = 1.5$, and in blue, assuming $f_{h_z}$ equal to the value which produces the best agreement between the DMSD  and SMSD. The best-fit value of $f_{h_z}$, as well as the associated value of $\mathcal{F}_b^{2.2h_R}$, is given for each of the three galaxies in the legend and in Table \ref{tab:diskmass_galaxy_data}.}
    \label{fig:dms_compare}
\end{figure*}

Using the PCA fitting method described in \citetalias{pace_19a_pca}, we have made resolved estimates of the stellar mass of the three galaxies in both the DiskMass and MaNGA observed samples. We transformed the resolved stellar mass into stellar mass surface density $\Sigma^*$ by noting the solid angle and the surface area at the fiducial redshift subtended by each MaNGA spaxel, and then deprojecting to face-on using inclinations derived in \citet{diskmass_vi}, a modest ($\sim 20\%$) correction. Though inclinations are provided in MaNGA for these galaxies, we have elected to use the DMS values, to better ensure like radii are compared to like radii. We have elected to show the derived stellar (dynamical) mass surface density for the PCA (DMS) data, since $\Sigma^*$ is independent of Hubble parameter $h$ for PCA/SPS (both area and luminosity have a $h^{-2}$ dependence). The $h$-corrections to the DMS dynamical-masses are small ($\sim 0.02~{\rm dex}$), but have been made. Figure \ref{fig:dms_compare} compares the radially-averaged total-dynamical and resolved SPS estimates of mass surface density, and we reproduce in Table \ref{tab:diskmass_galaxy_data} the relevant rows of \citet[][Table 4]{diskmass_vii}, which summarizes the kinematic decompositions of the two galaxies.

\begin{table*}
    \centering
    \begin{tabular}{||c|c||c|c|c|c|c||} \hline
        UGC & \texttt{plateifu} & $\mathcal{F}_b^{2.2h_R}$ (DMS) & $\hat{f}_{h_z}$ & $\mathcal{F}_b^{2.2h_R}$ ($\hat{f}_{h_z}$) \\
        (1) & (2) & (3) & (4) & (5) \\ \hline
        3997 & 8566-12705 & 0.48 & $0.90 \pm 0.12$ & 0.46 \\ \hline
        4107 & 8567-12701 & 0.56 & $1.83 \pm 0.15$ & 0.76 \\ \hline
        4368 & 8939-12704 & 0.72 & $1.68 \pm 0.10$ & 0.93 \\ \hline
    \end{tabular}
    \caption{Summary of comparisons between DMSD and SMSD, and the effect of a changing disk scale height on the estimated baryon fraction: column (3) ($\mathcal{F}_b^{2.2h_R}$) is the fractional contribution of the stellar disk to the rotation curve at 2.2 disk scale-lengths reported in \citet{diskmass_vii}. Columns (4) and (5) give $\hat{f}_{h_z}$ (the value of $f_{h_z}$ which gives the best galaxy-wide match between $\Sigma_{dyn}$ and $\Sigma^*$), and the associated value of $\mathcal{F}_b^{2.2h_R}$.}
    \label{tab:diskmass_galaxy_data}
\end{table*}

Given that the mass in stars heavily dominates the velocity structure of all of these galaxies, one would expect the dynamical-mass surface density (DMSD) would equal or only slightly exceed the stellar-mass surface density (SMSD), because of the mass contributed by gas, a bulge, or the dark matter halo---so, why are the dynamical and SPS estimates of mass-density systematically discrepant? This comparison is an apt illustration of the distinct systematics of stellar population synthesis and disk kinematic fitting.

\citet{aniyan_freeman_16, aniyan_freeman_18} respectively address issues with matching appropriate values of $h_z$ for a given measurement of $\sigma_z$ (i.e., the adopted scale-height must be that of a similar dynamical tracer), and the contribution of a dark matter halo to the stellar $\sigma_z$. Assuming disks are maximal places upper limits on the disk mass surface-density (and hence, upper limits on stellar mass-to-light ratio and lower limits on disk thickness). That is, accounting for dark matter in the dynamical models tends to lower the estimate of disk maximality \citep{hessman_17}. Lower limits on disk mass surface density (hence, lower limits on stellar mass-to-light ratio and upper limits on disk thickness) are reasonably bounded by the DMS measurements: the scale height of the stellar dynamical tracer is not larger than the observed broad-band disk thickness, and dark matter corrections are modest \citep{diskmass_iii, swaters_14}. However, since DMS measurements indicate that disks are substantially sub-maximal, there is room to consider adjustments to the adopted disk thickness to bring dynamical values closer in line with our SPS estimates. At issue is not the scaling relation adopted in \citet{diskmass_ii} since edge-on dynamical measurements where scale heights are directly accessible give similar results \citep{bershady_11}. As \citep{aniyan_freeman_16} posit, though, the scale height relevant for the stellar dynamical tracer could be systematically smaller than the measured broadband values.

As an illustration of the effects of scaling $h_z$, we have considered a galaxy-wide overestimate of $h_z$ by a factor $f_{h_z}$. In such a case, the stellar-mass surface density implied by a dynamical-mass surface density and baryon fraction (neglecting contributions from gas) is given by
\begin{equation}
    \Sigma^* \approx \Sigma_{dyn} f_{h_z}
\end{equation}
We first bin PCA-derived measurements of SMSD in radial bins which correspond to the DMS radial bins. We exclude radial bins interior to the bulge radius, and optimize to obtain the factor ($f_{h_z}$) most consistent with the galaxy-wide offset between SMSD and DMSD. We list this best-fit value in Table \ref{tab:diskmass_galaxy_data}, Column 4, along with the implied baryon-fraction associated with this change in $f_{h_z}$ ($\mathcal{F}_b^{2.2h_R}(f_{h_z}) \approx \mathcal{F}_b^{2.2h_R} \sqrt{f_{h_z}}$: Table \ref{tab:diskmass_galaxy_data}, Column 5). The uncertainty in $f_{h_z}$ is largely dictated by the intrinsic dispersion of SMSD in a given radial bin. Allowing for the possibility that the PCA masses are systematically high by about 0.1 dex could provide some relief from large values of $f_{h_z}$, but does not make dynamics and SPS consistent. This allowance also worsens the mismatch for UGC3997 (8566-12705), which already has $f_{h_z} < 1$. A more complete analysis, though, must re-propagate the dynamical effects of the gas, which has a different scale-height and must be considered separately (though UGC3997 and UGC4107 have such low gas mass surface densities in comparison to stellar that the additional correction should be quite small).

This brief exercise does not paint a clear picture of whether $h_z$ is significantly underestimated: for the case of UGC3997---by far the lowest-surface-brightness of the three---, $f_{h_z}$ is implied to be less than unity\footnote{In fact, for UGC3997, the inferred $\Sigma_{dyn}$ and $\Sigma^*$ profiles are quite different in slope, which implies that there might be another latent effect at play. A flared disk (i.e., one whose scale height increases with radius) is strongly disfavored, because its DMSD profile would have a steeper slope with respect to a given SMSD profile. This is the opposite effect from what we see for UGC3997.}, while for the remaining two galaxies, the favored values of $f_{h_z}$ are 1.83 and 1.68, somewhat less than the factor of two that \citet{aniyan_freeman_16} favors. While claims regarding specific values of $f_{h_z}$ are outside the scope of this paper, clearly modest corrections to $h_z$ bring $\Sigma_{dyn}$ into qualitative agreement with our SPS-derived values of $\Sigma^*$.

In \citetalias{pace_19a_pca}, we treated in some detail the systematics in stellar population fitting, and justified our various choices in parametrizing the SFH training data. However, we further expound here regarding two important and outstanding issues. First, our relatively burst-poor SFHs may be systematically too heavy \citep{gallazzi_bell_09}. If $h_z$ were more closely-constrained at the values found by the DiskMass Survey, allowing more stochastic variations in instantaneous SFR might be justified. Regardless, the above comparisons also justifies more detailed treatments of disk dynamics with dark matter \citep{hessman_17} and more accurate measurements of disk thickness and vertical mass-to-light ratio gradients \citep{asr_mab_13, asr_mab_14, eigenbrot_bershady_18}. 

Second, these results rely on the assumption of a radially-uniform IMF, which is seen to change both from galaxy to galaxy \citep[for a JAM-SPS modelling comparison, see ][]{li_17_radIMF} and within a single galaxy \citep{martin-navarro_15_radIMF, la-barbera_16_radIMF}. As \citet{li_17_radIMF} reports, a bottom-heavy (-light) IMF in the inner (outer) galaxy will bring about a more negative stellar mass-to-light ratio gradient as compared to dynamics. We see no evidence for such a discrepancy in Figure \ref{fig:dms_compare}, but the three particular galaxies examined here are likely not the best candidates for a search for such a trend; galaxies more widely sampled in radii may shed further light on this matter.

More careful integration of resolved SPS with Jeans and Schwarzschild modelling may in the future permit direct comparison between stellar and dynamical masses for elliptical galaxies \citep{li2019_manga_etg_densityslopes}. A simple comparison between the projected (plane-of-sky) SMSD and DMSD is not by itself informative for several reasons: first, the majority of such studies assume a radially-constant stellar mass-to-light ratio, which is at minimum inconsistent with this work's assumptions; second, the integration of the (three-dimensional) posterior dark matter density profile along lines-of-sight requires adopting a cutoff radius for the dark matter halo \citetext{Zhu priv. comm.}, a radius which will be larger than the field of view subtended by a MaNGA IFU, since galaxies are only sampled to 2.5 $R_e$ \citet{manga_sample_wake_17}; and third, while the volume-density profile will be dominated in the centers of ellipticals by the stars, it will appear in projection that all parts of an elliptical galaxy are dark matter-dominated, due to the long ``column" of dark matter-dominated galaxy outskirts along each line-of-sight. For this reason, we do not ourselves make any comparisons using elliptical galaxies.

\section{A catalog of aperture-corrected total galaxy stellar masses}
\label{sec:mstar_catalog}

In addition to resolved stellar mass maps, we provide very approximate aperture-corrections to estimate total galaxy stellar mass. The basic task is to calculate the approximate amount of stellar luminosity outside the grasp of the IFU, and multiply it by a reasonable guess for the stellar mass-to-light ratio. This involves accounting for both regions outside the IFU proper and regions with known bad data quality. The latter case is handled first, for which cases the adopted mass-to-light ratio is the median of the eight nearest unmasked spaxels' values.

The limited spatial coverage of the MaNGA IFU at radii larger than 1.5 $R_e$ (2.5 $R_e$) in the Primary+ (Secondary) Sample will cause the outskirts of galaxies to remain unsampled. \citet{gonzalez-delgado_14} proposed a simple solution for CALIFA data, where additional mass is added by adopting the average mass-to-light ratio between 1.5 and 2.5 $R_e$ as a fiducial mass-to-light ratio of the galaxy outskirts: since stellar mass-to-light ratio is observed to not change significantly in the outermost regions of disks, this seems a reasonable choice.

\begin{figure}
    \centering
    \includegraphics[width=\columnwidth]{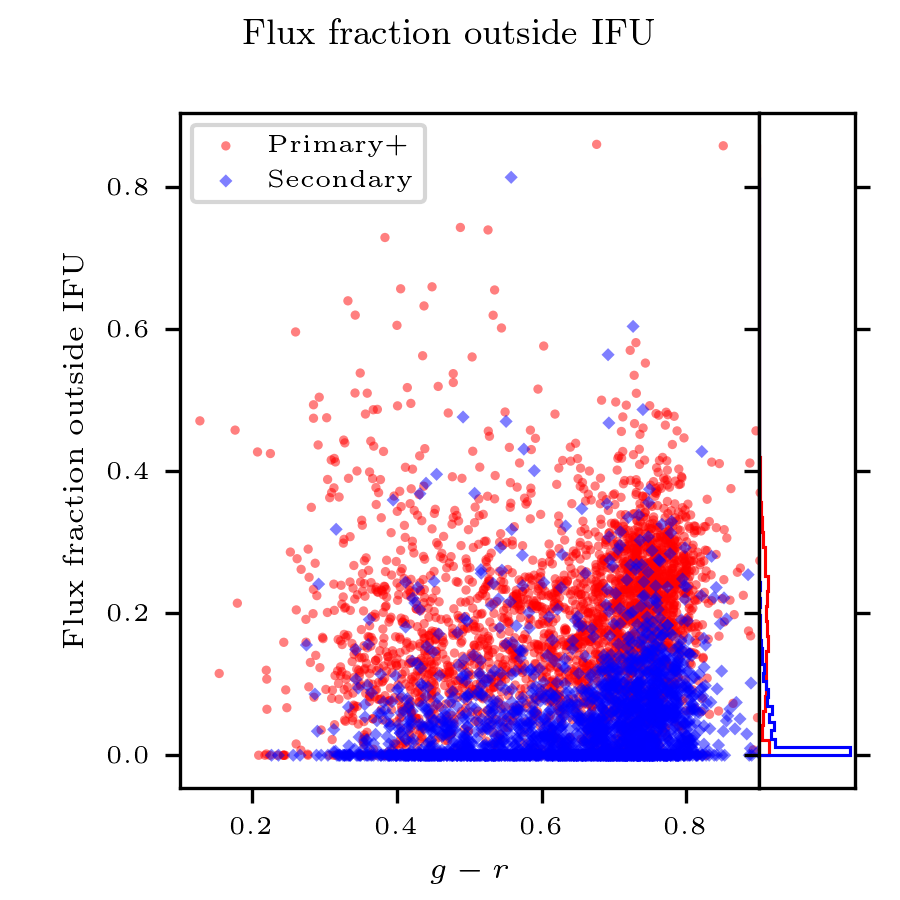}
    \caption{The fraction of $i$-band flux falling outside the grasp of the MaNGA IFU, plotted against $g-r$ color, and separated by whether the galaxy is in the Primary+ (red circle) or Secondary (blue diamond) sample. In the right panel is shown the distribution of flux fraction outside the IFU, separated by the sample designation (Primary+ or Secondary).}
    \label{fig:flux_outside_ifu}
\end{figure}

In order to choose the aperture-correction method, we first evaluate how much of the galaxy's luminosity falls outside the MaNGA IFU. In Figure \ref{fig:flux_outside_ifu}, we separate galaxies by sample (Primary+ and Secondary), and plot the flux deficit between the IFU-summed PCA-reconstructed stellar continuum bandpass flux and the K-corrected bandpass flux from the NASA-Sloan Atlas \citep{blanton_roweis_07}. 80\% of galaxies lose less than 22\% of their integrated, $i$-band flux to aperture effects, and the vast majority of those are in the Primary+ sample. Many such galaxies have a foreground star just outside the IFU, which (if improperly isolated in the photometry) might bias the flux fraction outside the IFU to a higher value.

Ultimately, we must choose a constant mass-to-light ratio for the galaxy outskirts, and we consider two methods:

\begin{enumerate}
    \item \textbf{``CMLR method":} compute a color or the missing flux, and convert it into a stellar mass-to-light ratio using a CMLR (such as the one found in \citealt{pace_19a_pca}).
    \item \textbf{``Ring method":} adopt as the stellar mass-to-light ratio for the aperture-correction the median value in a ring in the outermost accessible part of the galaxy. For high-inclination galaxies ($\frac{b}{a} < 0.5$), we average over spaxels within 30 degrees of the major axis, and in the outermost 0.5 $R_e$ of the galaxy with well-determined stellar mass-to-light ratios. By restricting our sampling to along or nearby the major axis, we better ensure that we are averaging over like radii (in contrast, including spaxels along the minor axis probes a range of radii even along a single line of sight). For galaxies with $\frac{b}{a} > 0.5$, we average over the same 0.5 $R_e$ ellipse, but with no restrictions on azimuthal angle relative to the major axis.
\end{enumerate}

\begin{figure}
    \centering
    \includegraphics[width=\columnwidth]{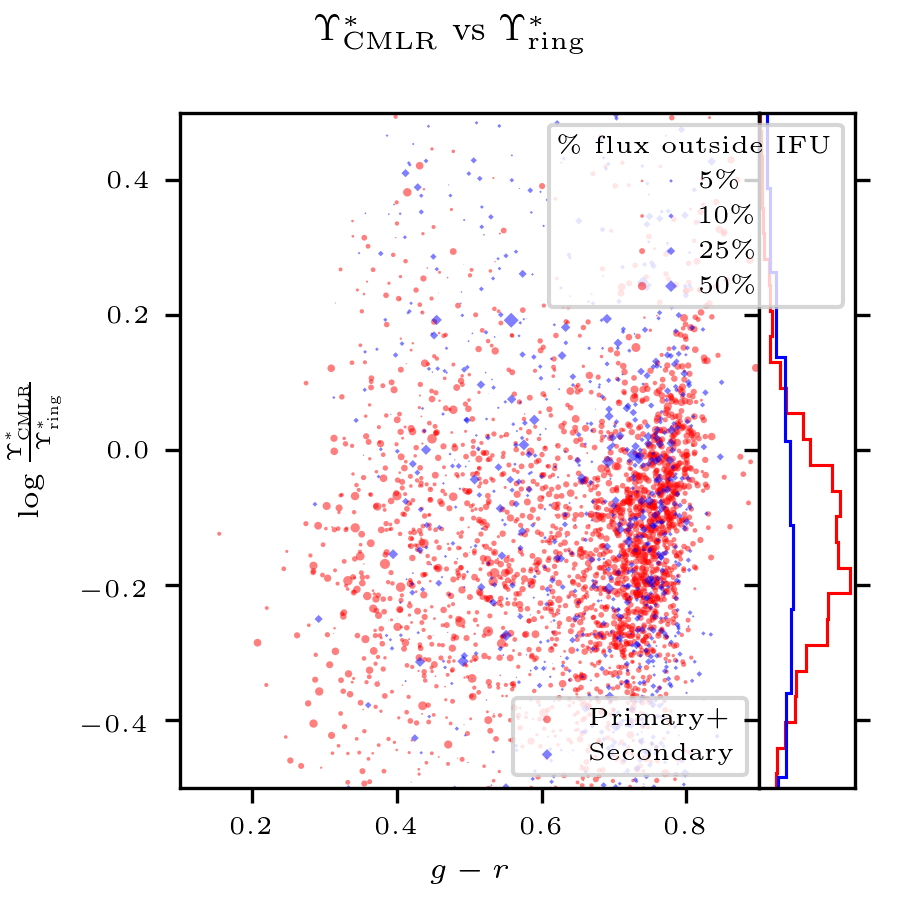}
    \caption{The difference in inferred stellar mass-to-light ratio for the galaxy outskirts between the ring method and the CMLR calibration. Point color and shape are the same as Figure \ref{fig:flux_outside_ifu}, and points are sized according to the fraction of total luminosity outside the IFU. Galaxies whose reconstructed-$i$-band flux equals or exceeds the K-corrected NSA $i$-band flux are not displayed. Shown in the right panel is the distribution of $\log \frac{\Upsilon^*_{\rm CMLR}}{\Upsilon^*_{\rm ring}}$, separated by sample (Primary or Secondary).}
    \label{fig:outer_ml}
\end{figure}

Though the difference between the two methods has little discernable systematic on average, for an individual galaxy, $\frac{\Upsilon^*_{\rm CMLR}}{\Upsilon^*_{\rm ring}}$ exhibits some extreme values at extreme colors of residual flux. Granted, these seem to primarily be galaxies with little residual flux at all (see Figure \ref{fig:outer_ml}), and those cases ought to be quite susceptible to measurement uncertainties.

Figure \ref{fig:outer_ml} shows the dependence of the difference between ``ring" and ``CMLR" stellar-mass aperture-correction on optical ($g-r$) color and sample (the Primary Sample has coverage out to at least 1.5 $R_e$, and the Secondary out to at least 2.5 $R_e$). Point size illustrates the flux fraction outside the IFU (as in Figure \ref{fig:flux_outside_ifu}). The secondary sample seems to experience greater scatter in inferred stellar mass-to-light for the galaxy outskirts, but this results in very small luminosity-scalings for these corrections. In other words, when bandpass fluxes derived from MaNGA spectroscopy are very close to (but still less than) the catalog flux, a small discrepancy in one photometric band (relative to another) will produce an extreme color, which would imply an extreme mass-to-light ratio.

\begin{figure}
    \centering
    \includegraphics[width=\columnwidth]{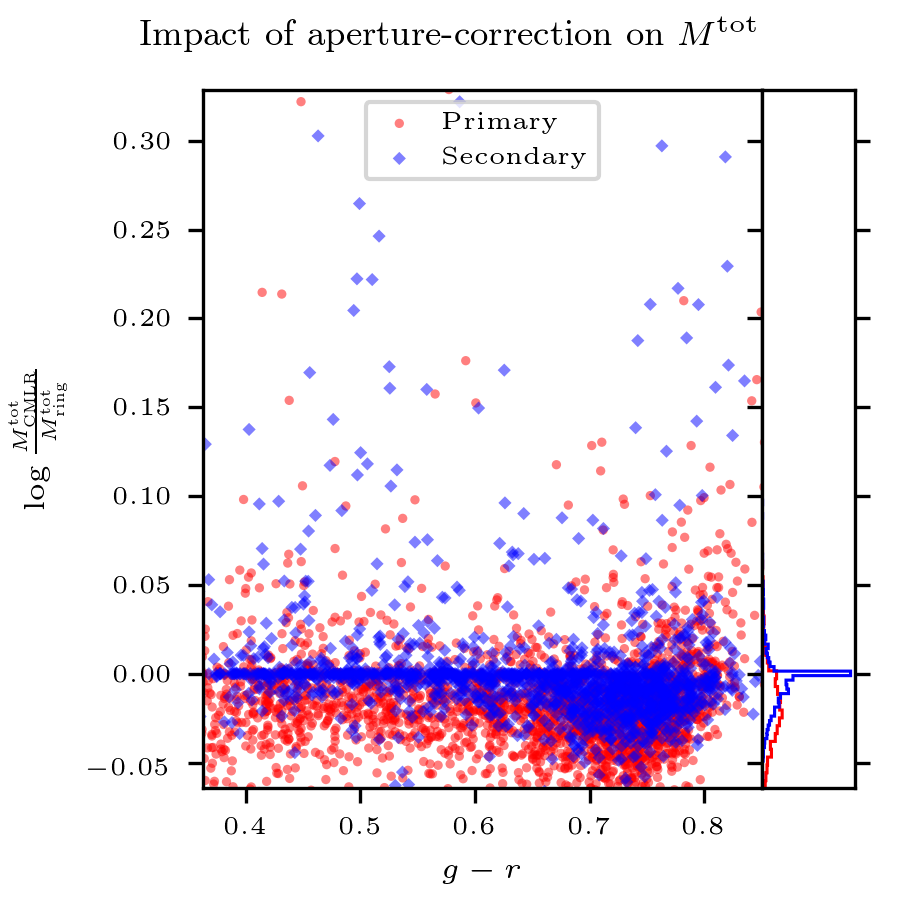}
    \caption{The difference in \emph{total stellar mass} induced by a choice of stellar mass-to-light ratio for the outskirts (ring model or CMLR), versus $g-r$ color. Points are colored according to their sample, as Figure \ref{fig:flux_outside_ifu}. Galaxies fully covered by the IFU (i.e., with no missing $i$-band flux) have their values of $\log \frac{M^{\rm tot}_{\rm CMLR}}{M^{\rm tot}_{\rm ring}}$ set to zero. In the right panel is the distribution of $\log \frac{M^{\rm tot}_{\rm CMLR}}{M^{\rm tot}_{\rm ring}}$, separated by sample (Primary or Secondary).}
    \label{fig:mtot_compare_cmlr_ring}
\end{figure}

Given the potential for color-correlated systematics seen above, we also evaluate the impact this has on total galaxy stellar mass: after all, if the flux outside the IFU is very small, a disagreement in inferred mass-to-light ratio is unimportant. Indeed, Figure \ref{fig:mtot_compare_cmlr_ring} shows that the difference in the \emph{total stellar mass} induced by changing the aperture-correction method is much less than 0.1 dex. However, the locus of points at a fixed color indicates that the ``ring" method does produce very slightly higher masses. Furthermore, as expected, the Primary sample (being sampled to a smaller radius, on average) is aperture-corrected more severely, so the differences between the two methods is more prominent. In addition, we find that neither method produces noticeably more extreme color-correlated systematics with respect to previous photometrically-obtained stellar masses (see Section \ref{sec:photometry}). 

We recommend adopting the ``CMLR" method: while we expect that the resulting mass-to-light ratios have systematics at the 0.1 dex level or below, this approach implicitly compensates for radial mass-to-light ratio gradients \citep{tortora_napolitano_10_spgrad, tortora_napolitano_11, boardman_weijmans_vdB_17} by varying the fiducial stellar mass-to-light ratio based on the actual ``missing flux". Therefore, we avoid (for example) chronically overestimating the mass in the outer regions of a galaxy with a negative stellar mass-to-light ratio gradient. That said, one could make the case for adopting the ``ring" correction for the Secondary sample, since the discrepancy in total mass should be reduced relative to the Primary sample, assuming a negative mass-to-light ratio gradient. Put another way, any adopted aperture-corrections are only as good as the imaging and spectroscopically-synthesized photometry used to estimate the bandpass flux deficits. In our catalog of aperture-corrections, we will include both potential mass aperture-corrections, with the caveat the the ``CMLR" method is likely safer in the general case.

\subsection{Tests of aperture-correction against resolved photometry}
\label{subsec:aper_photometry}
The above comparisons between the two proposed aperture-correction methods indicate that in most cases, it is preferable to adopt a fiducial mass-to-light ratio for the regions of the galaxy outside the IFU based on the color of the residual flux in the $g$ and $r$ bands and a CMLR. It would be desirable, however, to check the stellar masses which would result from applying a CMLR to resolved photometry of MaNGA galaxies against the spectroscopically-derived values. This test will shed light on the possibility of bias in the aperture-corrections. To that end, we exploit the legacy multi-band pre-imaging which exists for all MaNGA galaxies \citep{sdss_summary}.

\begin{figure*}
    \centering
    \includegraphics[width=\textwidth]{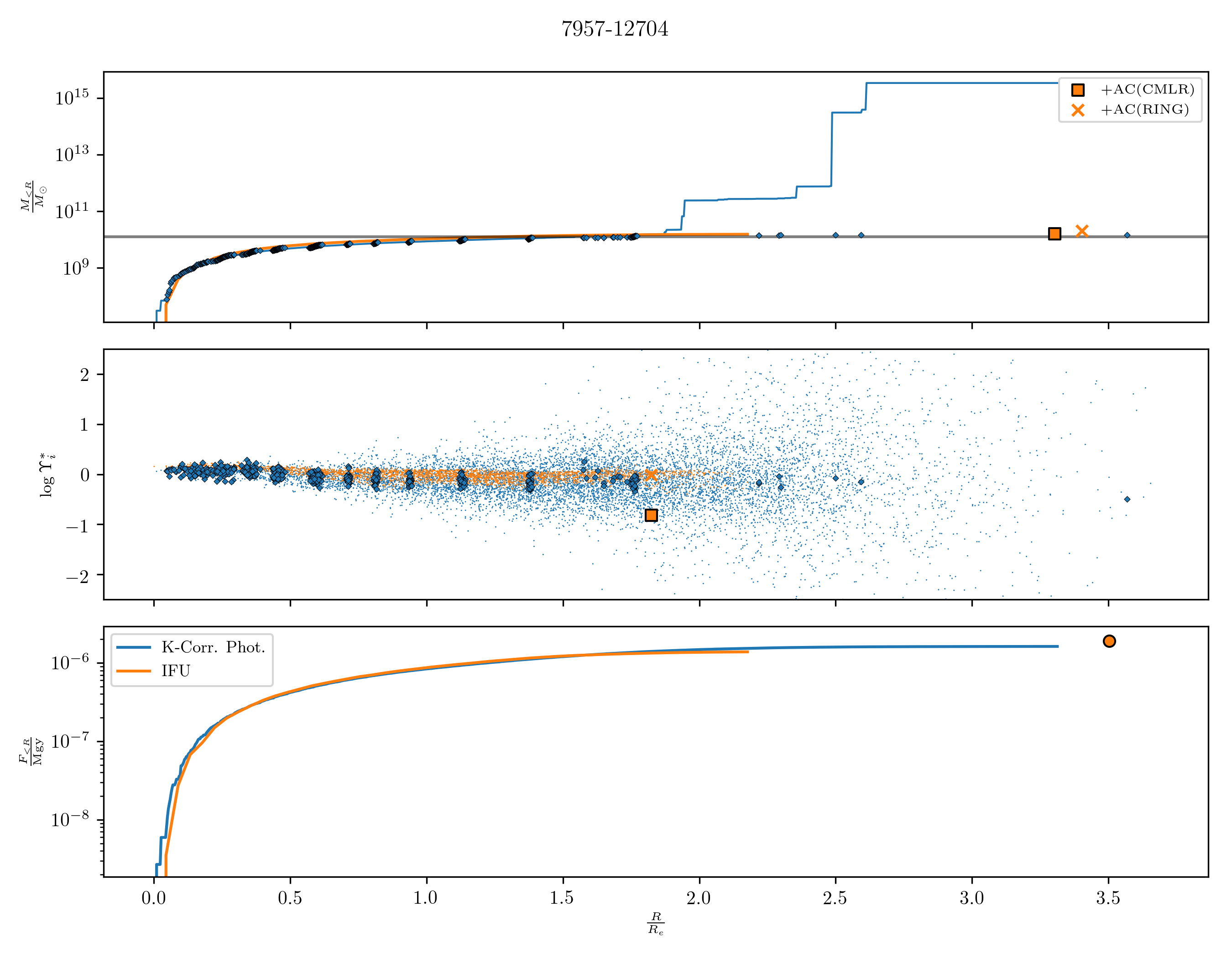}
    \caption{An example spectroscopic-photometric aperture ``curve-of-growth" comparison figure. \textit{Top panel:} Cumulative stellar-mass enclosed within ellipses of increasing radius (units of $R_e$), found using photmetry and a CMLR (blue line) with spectroscopic PCA methods (orange line). For the photometry, low signal-to-noise (found at high radii) biases mass-to-light high and introduces bias (the blue curve increases without bound). Blue diamonds show cumulative masses from radially- and azimuthally-binned photometry with the fiducial CMLR. The orange $\blacksquare$ and \texttt{X} refer to the total stellar-mass of the galaxy after aperture-correction with respectively the ``CMLR" and ``ring" methods. The horizontal, black, dashed line denotes the stellar-mass from the NSA. \textit{Middle panel:} radial plot of $i$-band mass-to-light ratio: blue dots denote pixels from preimaging, orange dots spaxels from the MaNGA datacube, and blue diamonds denote radiall-azimuthally binned photometry; and the single, orange $\blacksquare$ and \texttt{X} denote the mass-to-light ratio that would respectively result from the ``CMLR" and ``ring" aperture-corrections. \textit{Bottom panel:} As top panel, but radially-summed fluxes.}
    \label{fig:specphot_compare}
\end{figure*}

Figure \ref{fig:specphot_compare} offers (for a single galaxy) a comparison between the stellar masses \& stellar mass-to-light ratios resulting from the PCA analysis, and those emerging from simply applying a fiducial CMLR (from \citetalias{pace_19a_pca}) to individual pixels from the SDSS-I preimaging cutouts \citep{sdss_summary}. In the case shown (and in most cases), the cumulatively-radially-summed mass yielded by combining single preimaging pixels with a CMLR gives an unphysically-high mass. Since the CMLR gives the logarithm of the stellar mass-to-light ratio, low signal-to-noise pixels which are anomalously red produce anomalously-high stellar mass-to-light ratios (and thus, huge masses); but anomalously-blue pixels produce next to no mass at all \emph(but not negative mass). The net effect is for total mass to increase radially without bound. As a counterpoint to this, we show a similar cumulative mass profile which results from coadding nearby preimaging pixels to reach some target signal-to-noise ratio in the $i$-band, and alleviating this issue. At the same time, the overall slope of the radial mass-to-light ratio profile is preserved relative to the PCA. Also shown are the aperture-corrections resulting from the ``CMLR" and ``ring" methods.

In the case shown, and indeed for most galaxies, the adopted mass-to-light ratio for the ``ring" aperture-correction (orange \texttt{X} in Figure \ref{fig:specphot_compare}) is sizably greater than the ``CMLR" value (orange $\blacksquare$). Once again, we believe this has \emph{some} basis in reality, because optical colors do overall trend blueward with increasing radius inside a galaxy (i.e., negative stellar mass-to-light ratios). That said, in many cases, the difference between the two mass-to-light ratios is larger than might be expected. 

In all cases, the flux enclosed within small annuli is larger for the preimaging data than for the photometry reconstructed from the MaNGA spaxels: this is because the preimaging pixels are a factor of several smaller than the MaNGA spaxels (so, flux from the preimaging will be counted at a smaller radius than flux from the MaNGA datacube). The effect becomes insignificant at higher radii, because galaxy surface brightness gradients get less steep with radius. Additionally, in some rare cases (less than 5\% of the time), the total enclosed flux at the largest radius in the IFU is smaller than the enclosed flux for the photometry by approximately 10\% at the same radius. This does not seem to depend on redshift or color, nor does it seem to correspond with the presence of foreground stars.

Knowing what we do about galaxy stellar mass-to-light ratio gradients, it is reasonable to conclude that the ``ring" method likely provides a relatively strong upper-limit on the stellar mass outside the IFU, especially for galaxies in the Primary sample. As shown in Figure \ref{fig:mtot_compare_cmlr_ring}, the logarithmic mass differential induced by changing the aperture-correction method is not as large in the Secondary sample as in the Primary sample, though the effect is small (less than 0.1 dex) across the board. Galaxies sampled to a higher effective radius on average will have in their outermost spaxels a lower mass-to-light ratio. Furthermore, though the individual photometric residuals used by the CMLR might be more uncertain, the overall correction will also be smaller on average. Ultimately, for the Secondary sample, both aperture-corrections are likely equally usable.

\subsection{Integrated masses: comparison to NASA-Sloan Atlas \& JHU-MPA Catalog}
\label{sec:photometry}

As a point of comparison with the systematics we have uncovered and discussed so far, we also attempt to characterize the differences between the IFU-summed and aperture-corrected stellar masses from PCA and those obtained by analyzing photometric data from SDSS \citep{sdss_summary}: we consider simultaneously the NASA-Sloan Atlas, which fits a set of SFH templates to elliptical-Petrosian photometric fits, in order to obtain K-corrected rest-frame magnitudes and stellar-masses \citep{blanton_11_nsa, blanton_roweis_07_software}, and the JHU-MPA catalog, which uses \textit{ugriz} photometry to estimate dust attenuation, mass-to-light ratio, and broadband, K-corrected luminosities \citep{kauffmann_heckman_white_03, salim_rich_charlot_07, sdss_dr8_2011}. 

\begin{figure}
    \centering
    \includegraphics[width=\columnwidth]{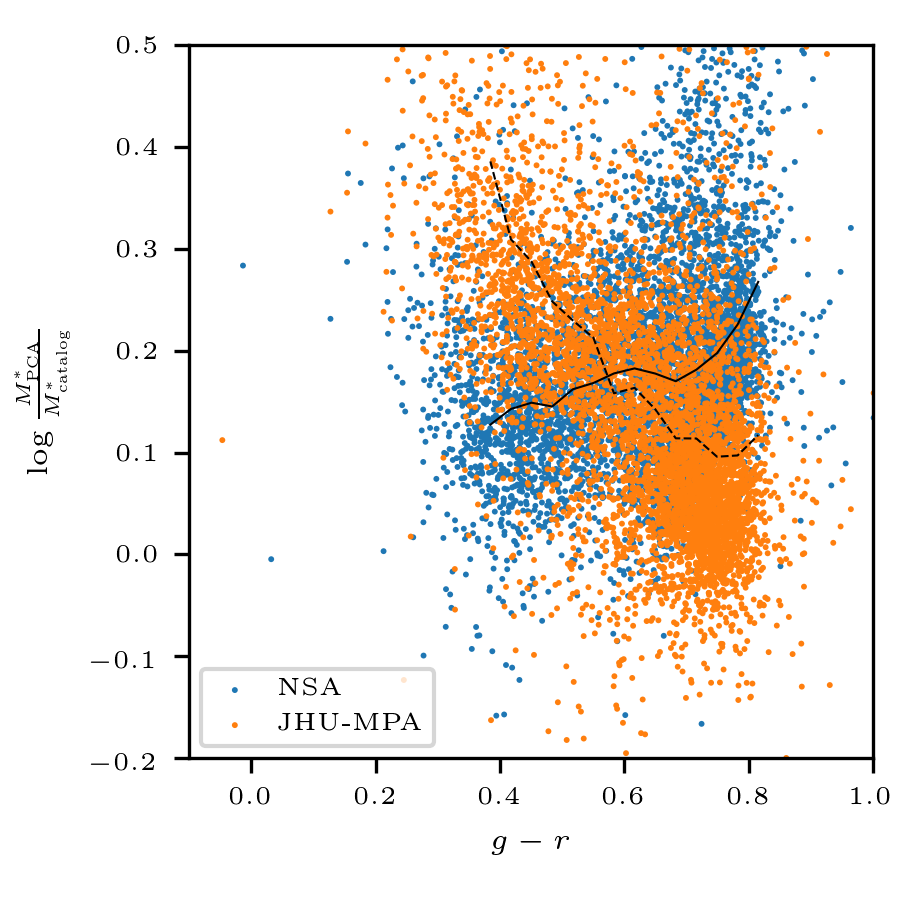}
    \caption{For \nrungalaxies galaxies, the difference between the stellar masses from this study's PCA analysis and those from the NSA (blue points, and a locally-weighted regression as a solid, black line); and the difference between the stellar masses from this study's PCA analysis and those from the JHU-MPA galaxy catalog (orange points, and a locally-weighted regression as a dashed, black line). Both NSA and JHU-MPA masses have been corrected to a Kroupa IMF.}
    \label{fig:dMasses}
\end{figure}

In Figure \ref{fig:dMasses}, we compare the PCA-derived stellar mass estimates from this work to those reported in the NSA and the JHU-MPA catalog (both catalogs have been corrected to $h=.693$, and to a Kroupa IMF). We note that the PCA masses are almost universally the highest, exceeding both libraries by approximately 0.1--0.15 dex at red colors (this discrepancy increases for the JHU-MPA catalog at bluer colors, to a maximum of 0.3--0.4 dex). Scatter about the mean deviation is approximately 0.2 dex across all colors, and 0.1 dex at fixed color. Possible responsible effects include stellar libraries (even in old SSPs, \logml{i} can vary systematically by $\sim 0.05$ dex) and simple differences in SFHs. To wit, the SFHs employed in this work are much more continuous than those adopted in previous work: the majority of the training data used in \citet{blanton_roweis_07} were SSPs, and \citet{kauffmann_heckman_white_03} permitted a burst on average once per 2 Gyr after formation. Finally, it should perhaps be encouraging that at the redder colors ($0.5 \le (g-r) \le 0.8$) corresponding to intermediate-age stellar populations, the mass discrepancies lessen.

The systematics shown above are not unique to PCA mass measurements, and will likely be found between any combination of measurements, be they photometric or spectroscopic. For example, $\log \frac{M^*_{\rm JHU/MPA}}{M^*_{\rm NSA}}$ varies from -0.2 dex at $g-r \sim 0.3$ to 0.2 dex at $g-r \sim 0.8$. Furthermore, even changing the method of measuring photometry can induce systematics, as well: The NSA measures magnitudes in two different ways, fitting a S\'{e}rsic profile and fitting an elliptical-Petrosian aperture. We elected to compare to masses derived from elliptical-Petrosian photometric estimates, because they are less prone to catastrophically poor fits \citep[see][for more discussion]{manga_sample_wake_17}. That said, elliptical-Petrosian fits are bluer by $\Delta (g-r) ~ .046 \pm .088$ for the full MaNGA sample, which induces a 0.1 $\rm dex$ difference in the NSA stellar masses.

Even if these discrepancies do indicate a real overestimate of stellar-mass by the PCA technique, the overall conclusions reached by comparing to the DiskMass dynamical masses do not change: the implied scale-height correction would decrease in magnitude, but would not be consistent with unity. Furthermore, this brings about the \emph{opposite} problem in the case of UGC3997, whose best-fit correction factor $f_{h_z}$ is \emph{less than one} using the fiducial stellar masses (so, simply applying the 0.1 dex systematic as a further ``correction factor" would bring the two measurements \emph{more out of line}.) However, as noted above, the difference in slope between the masses from SPS and dynamics implies that some other effect is at play.

\subsection{Are stellar masses derived from galaxy-coadded spectroscopy reliable?}
\label{subsec:resolution_effects}

We also aim to test the scenario explored by \citet{zibetti_2009}, where integrated galaxy colors were found to produce systematically low mass-to-light ratios, due to the effects of attenuation being incompletely captured. Any observations at coarser spatial resolution than resolved stellar populations could, in principle, suffer from similar biases.

\begin{figure}
    \centering
    \includegraphics[width=\columnwidth]{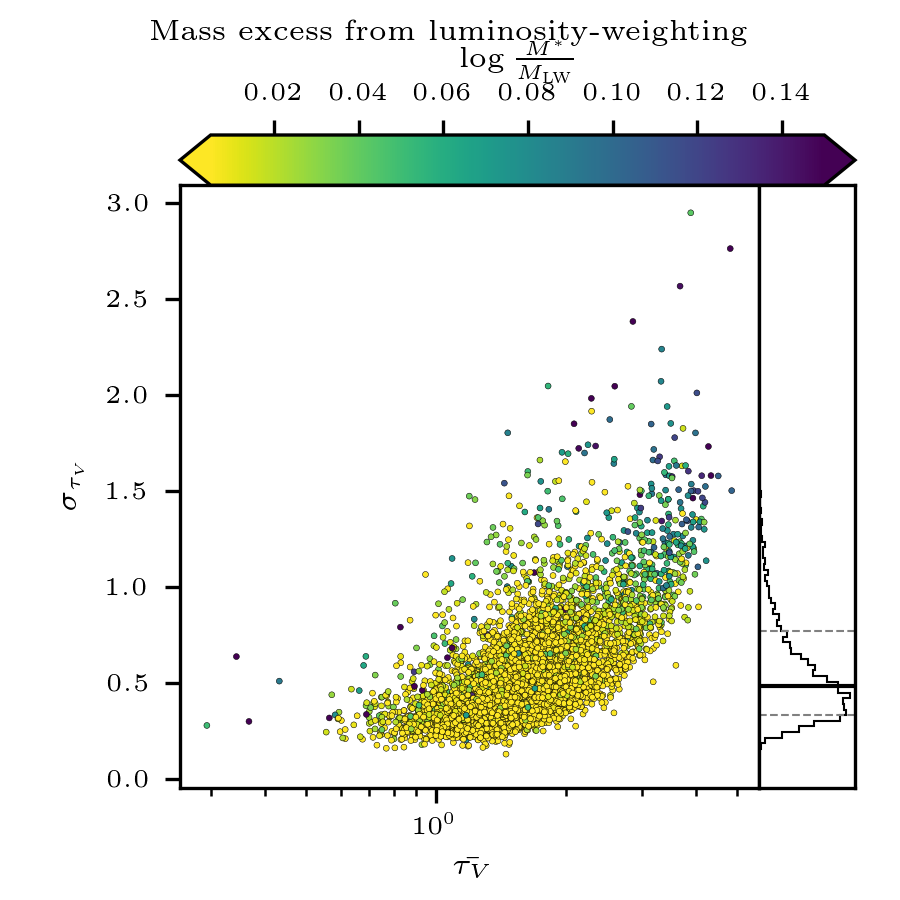}
    \caption{The logarithmic mass-deficit $\log{ \frac{M^*}{M^*_{\rm LW}} }$ induced by luminosity-weighting the spaxel-resolved stellar mass-to-light ratios (point color), plotted with respect to spaxel-stellar-mass-weighted dispersion in inferred $V$-band optical depth $\sigma_{\tau_V}$ and average inferred spaxel-stellar-mass-weighted $V$-band optical depth $\bar{\tau_V}$. The distribution of $\sigma_{\tau_V}$ is shown in the right-hand panel, and the 16$^{\rm th}$, 50$^{\rm th}$, and 84$^{\rm th}$ percentiles are shown as horizontal lines (dashed, solid, and dashed, respectively). The galaxies with the highest mass deficit have the highest degree of dispersion in local-stellar-mass-weighted $\tau_V$, even at fixed average $\tau_V$.}
    \label{fig:meantauV_dMglobloc_stdtauV}
\end{figure}

\begin{figure}
    \centering
    \includegraphics[width=\columnwidth]{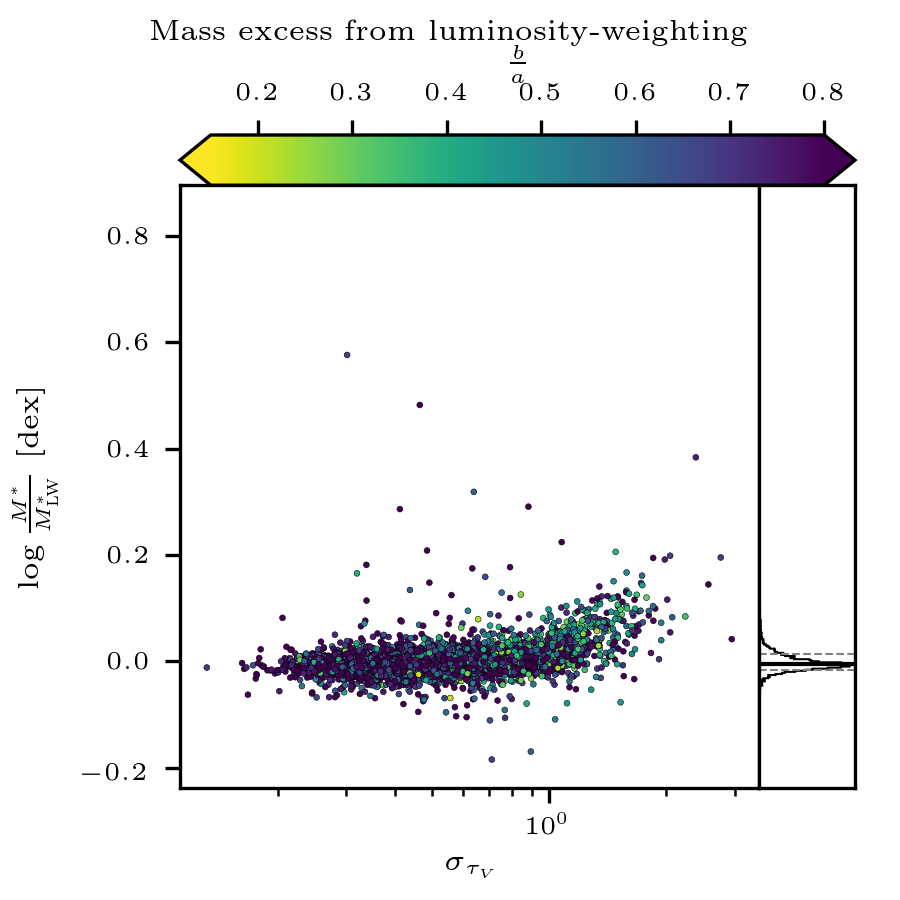}
    \caption{The logarithmic luminosity-weighting-induced mass deficit plotted against $\sigma_{\tau_V}$, and colored by NSA axis ratio. In the right-hand panel is shown the distribution of $\log \frac{M^*}{M^*_{\rm LW}}$.}
    \label{fig:meantauV_dMglobloc_ba}
\end{figure}

Here, we evaluate whether a spectrum for an entire galaxy (which will be dominated by light from young or relatively unattenuated stellar populations) will predict the same total galaxy stellar mass as a galaxy-coadded spectrum. In other words, we test the importance of implicit luminosity-weighting for spectra. To do so, we begin with a resolved map of a galaxy's stellar mass: we compare the sum over all resolved masses with the stellar mass obtained by flux-weighting the corresponding, spatially-resolved mass-to-light ratios\footnote{Spaxels with fitting errors, bad PDF population, or problematic spectroscopic data are excluded. In Section \ref{sec:mstar_catalog}, such spaxels were replaced by the median of their neghbors, but here we eliminate them altogether, so that all spaxels with good data are considered only once.}, and multiplying by the total luminosity in the IFU. We first attempt to explain the difference between the spaxel-summed mass ($M^*$) and the mass obtained from the luminosity-weighted mass-to-light ratios ($M_{\rm LW}$) by measuring how unequal the inferred dust attenuation is across the sampled area of the galaxy (represented by the standard-deviation of the distribution of local-stellar-mass-weighted optical depth, $\sigma_{\tau_V}$---Figure \ref{fig:meantauV_dMglobloc_stdtauV}). As more inclined galaxies are expected to exhibit more-strongly-attenuated lines-of-sight from our perspective, we also show a similar plot colored by elliptical axis ratio (Figure \ref{fig:meantauV_dMglobloc_ba}).

\begin{figure}
    \centering
    \includegraphics[width=\columnwidth]{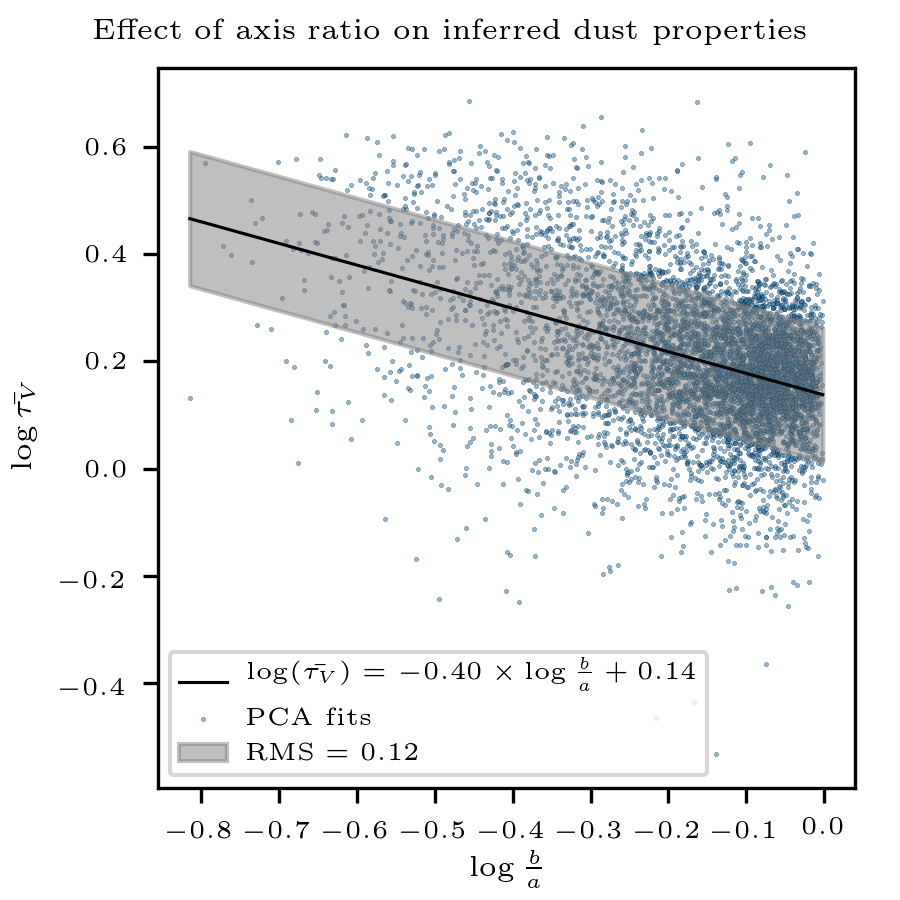}
    \caption{The correlation of spaxel-stellar-mass-weighted attenuation $\bar{\tau_V}$ on elliptical-Petrosian axis ratio $\frac{b}{a}$, both on a log scale. Blue points are individual galaxies analyzed using PCA, the black line is the least-squares fit, and the gray band is the RMS of the residuals.}
    \label{fig:meantauV_ba}
\end{figure}

We note that $\bar{\tau_V}$ correlates, as expected, with major-minor axis ratio $\frac{b}{a}$ (a proxy for inclination), with a power-law exponent of $\sim 0.40$ and a scatter of $\sim 0.12 ~ {\rm dex}$ (Figure \ref{fig:meantauV_ba}). This indicates that the PCA-based estimates of optical depth provide a reasonable indication of the influence of dust. The scatter at fixed axis ratio is considerable, though ($\sim 0.12 {\rm dex}$), which hints that simply adopting the axis ratio as a proxy for attenuation could neglect important galaxy-to-galaxy differences.

\begin{figure}
    \centering
    \includegraphics[width=\columnwidth]{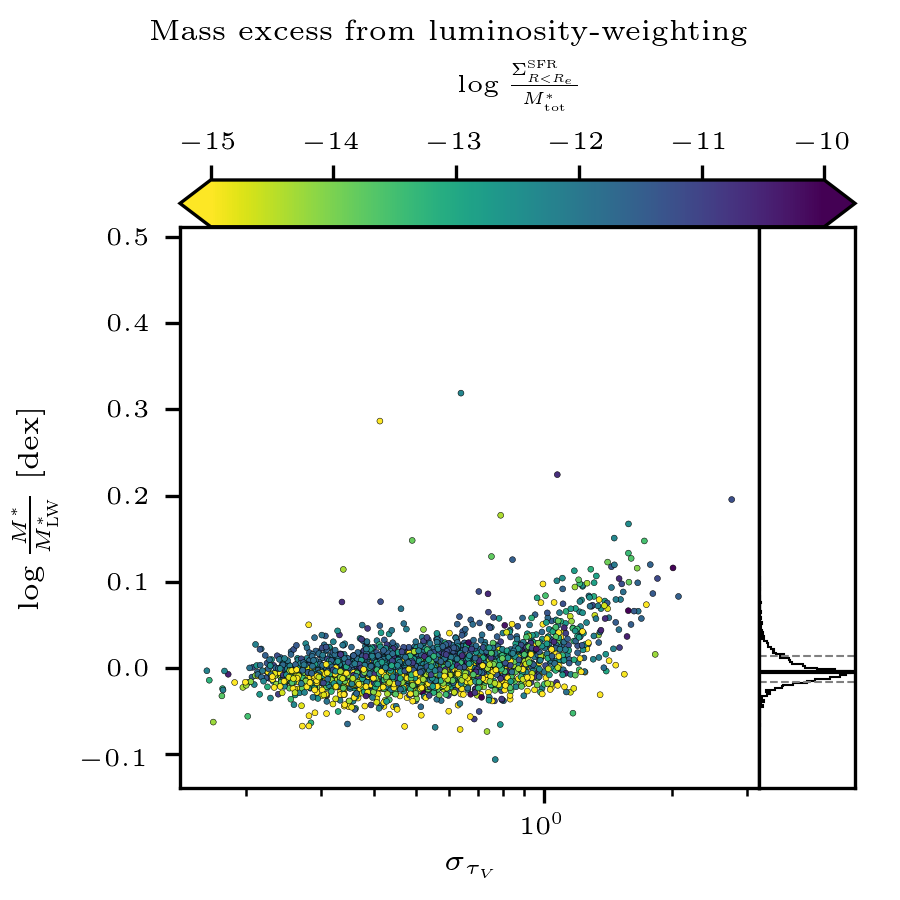}
    \caption{As Figure \ref{fig:meantauV_dMglobloc_stdtauV}, except colored by $\eta_{R<R_e} = \log \frac{\Sigma^{\rm SFR}_{R<R_e}}{M^*_{\rm tot}} f_{\rm corr}$, the specific star formation rate surface density within 1 $R_e$.}
    \label{fig:stdtauV_dMglobloc_ssfrsd}
\end{figure}

But, there could be additional important factors at play besides simply attenuation: \citet{sorba_sawicki_15} noted that, in the case of photometric stellar-mass estimates, $\log{M^*} - \log{M^*_{\rm LW}}$ was largest in galaxies with large values of sSFR, supporting the ``outshining" hypothesis. We find the total SFR in the innermost $R_e$ by selecting spaxels with emission-line ratios indicating ionization dominated by recent star-formation---i.e., lying below both \citet{kauffmann_03_agn} and \citet{kewley_dopita_01} limits---, co-adding their (dust-corrected) H$\alpha$ luminosities, and assuming the \citet{calzetti_13_sfr_indicators} SFR-H$\alpha$ calibration. H$\alpha$ luminosity has been corrected for dust attenuation to first-order, by adopting the stellar-mass-weighted mean attenuation as a fiducial V-band optical depth, and augmenting the H$\alpha$ luminosity by a factor of $f_{\rm corr} = {\rm exp}\left(\bar{\tau_V} \left(\frac{6564 \mbox{\AA}}{5500\mbox{\AA}}\right)^{-1.3}\right)$ \citep{charlot_fall_00}. We define a parameter $\eta_{R<R_e}$, intended to capture the majority of a galaxy's recent star formation, which divides the SFR surface density in the innermost 1 $R_e$ by the total, aperture-corrected stellar mass\footnote{We normalize by the total stellar mass to account for trends in star-formation rate surface density with total galaxy stellar mass.}:
\begin{equation}
    \eta_{R<R_e} = \frac{\Sigma^{\rm SFR}_{R<R_e}}{M^*_{\rm tot}} f_{\rm corr}
\end{equation}
and we experiment in Figure \ref{fig:stdtauV_dMglobloc_ssfrsd} with obtaining a tighter correlation than is seen in Figure \ref{fig:meantauV_dMglobloc_stdtauV} by replacing $\bar{\tau_V}$ with $\eta_{R<R_e}$.

Though there are high-mass-deficit galaxies at low-to-intermediate $\sigma_{\tau_V}$ with high $\eta$, such points do not seem to be more extreme in their central SFR than low-mass-deficit counterparts at similar $\sigma_{\tau_V}$. This provides some indication that strong central star formation is a subdominant effect to non-uniform attenuation when luminosity-weighting is concerned. In other words, the simple ``outshining" scenario may present less of a problem for spectroscopic measurements of stellar mass. At any rate, both dust and mean-stellar-age considerations are somewhat less important than the ($\sim 0.1$ dex) effects which can emerge from the choice of aperture-correction (see Section \ref{sec:mstar_catalog}). 

We attempt here an illustration of the relative importance of $\sigma_{\tau_V}$, $\bar{\tau_V}$, and $\eta_{R < R_e}$ for luminosity-weighted stellar mass, in order to summarize our findings: we use a multivariate ordinary least-squares fit of $\log \frac{M^*}{M^*_{\rm LW}}$ versus $\bar{\tau_V}$, $\sigma_{\tau_V}$, and $\log \eta_{R < R_e}$. We use the model
\begin{equation}
    \log \frac{M^*}{M^*_{\rm LW}} \sim \alpha + \beta_0 \bar{\tau_V} + \beta_1 \sigma_{\tau_V} + \beta_2 \log \eta_{R < R_e}
\end{equation}
and attempt to find values for $\alpha$, $\beta_0$, $\beta_1$, and $\beta_2$, the former of which sets a zeropoint for the luminosity-weighting-induced variation in stellar mass-to-light ratio in our sample, and the latter three of which respectively modulate the effects of average dust attenuation, differential dust attenuation, and specific star-formation rate surface density. In Table \ref{tab:m_lumwtd_regression_summary}, we summarize the OLS fit: of the three factors considered, $\sigma_{\tau_V}$ exhibits the strongest effects on luminosity-weighted, as indicated by the confidence interval for $\beta_1$ clearly excluding zero and the extremely low p-value. However, simply high, but uniform attenuation \emph{does not} induce an apparent mass-deficit after luminosity-weighting. Lastly, the regression indicates that $\log \eta_{R < R_e}$ can exert some effect, though small (an order of magnitude less severe than unequal attenuation). While we caution the reader against ascribing too much meaning to this regression (we do not include uncertainties on $\bar{\tau_V}$, $\sigma_{\tau_V}$, or $\log \eta_{R < R_e}$), it does provide us some insight that unequal attenuation is likely the dominant factor governing what modulates the effects of luminosity-weighting on stellar mass-to-light ratio at kiloparsec scales; and that an extremely large amount of recent star-formation could constitute a secondary influence. As this sample also contains a fair degree of contamination by galaxies with AGNs, foreground stars, or overlapping galaxies, such cases could be isolated in the future using visual classification.

\begin{table*}
    \centering
    \begin{tabular}{||c|c|c|c||} \hline
        Variable & Quantity modulated & CI (2.5--97.5) & p \\ \hline \hline
        $\alpha$ & zeropoint & [$1.38 \times 10^{-2}$, $2.61 \times 10^{-2}$] & $< 10^{-8}$ \\ \hline
        $\beta_0$ & $\bar{\tau_V}$ & [$-4.49 \times 10^{-3}$, $-1.38 \times 10^{-3}$] & $2.24 \times 10^{-4}$ \\ \hline
        $\beta_1$ & $\sigma_{\tau_V}$ & [$4.65 \times 10^{-2}$, $5.34 \times 10^{-2}$] & $< 10^{-8}$ \\ \hline
        $\beta_2$ & $\log \eta_{R < R_e}$ & [$2.83 \times 10^{-3}$, $3.69 \times 10^{-3}$] & $< 10^{-8}$ \\ \hline
    \end{tabular}
    \caption{Summary of the OLS fit of $\log \frac{M^*}{M^*_{\rm LW}}$ versus $\bar{\tau_V}$, $\sigma_{\tau_V}$, and $\log \eta_{R < R_e}$. For each of the ``slope parameters" $\beta$ and the ``zeropoint" $\alpha$, we give approximate confidence intervals and p-values.}
    \label{tab:m_lumwtd_regression_summary}
\end{table*}

\begin{figure*}
    \centering
    \includegraphics[width=\textwidth]{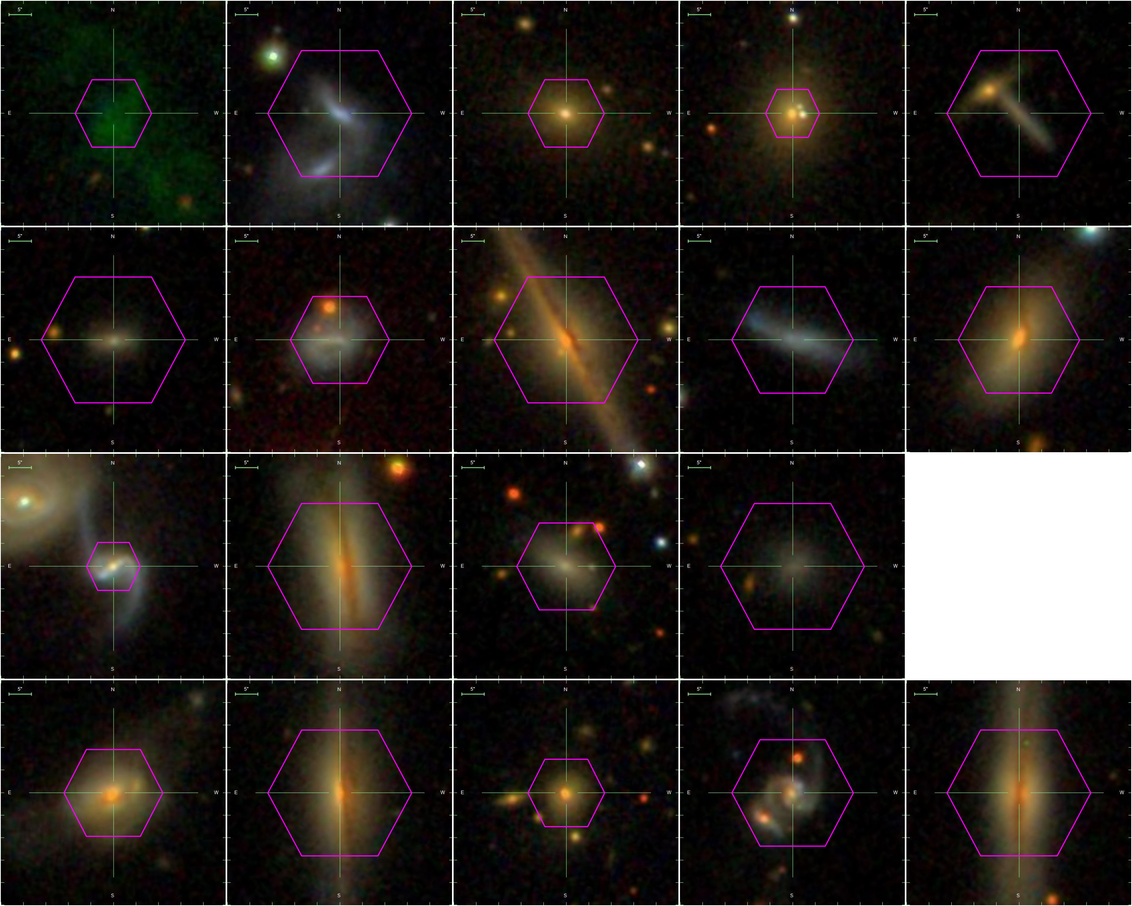}
    \caption{The twenty galaxies with the largest difference between $M^*$ and $M^*_{\rm LW}$. If read left-to-right and top-to-bottom, galaxies are in the same order as Table \ref{tab:rogues_gallery}. Galaxies with unavailable three-band image cutouts have been omitted.}
    \label{fig:rogues_gallery}
\end{figure*}

Finally, we produce a list of the twenty worst-offending differences between IFU-summed and luminosity-weighted stellar masses (Table \ref{tab:rogues_gallery}), and show images of each (Figure \ref{fig:rogues_gallery}). We exclude galaxies whose centers are contaminated by broad-line AGN emission: the vast majority of broad-line AGN have poor PCA fits; but the rare spaxel that is not flagged can cause trouble when compared to the rest of the galaxy, and areas with fainter (but still broad) AGN emission also normally have poor PCA fits.

\begin{table*}
    \centering
    \begin{tabular}{||c||c|p{4in}||} \hline
        MaNGA \texttt{plateifu} & $\log \frac{M^*}{M^*_{\rm LW}}$ [dex] & Description \\ \hline \hline
        8158-3703 & 0.576 & low-surface-brightness dwarf galaxy \\ \hline
        8139-12702 & 0.319 & two galaxies in IFU \\ \hline
        8999-3702 & 0.287 & possible poor stellar continuum fit, stellar velocity field \\ \hline
        8566-1901 & 0.224 & two foreground stars \\ \hline
        8249-12702 & 0.206 & two galaxies in IFU, different redshifts \\ \hline
        9869-12704 & 0.199 & foreground star \\ \hline
        8942-6102 & 0.192 & two foreground stars \\ \hline
        8080-12702 & 0.167 & edge-on galaxy with prominent dust lane \\ \hline
        8995-9102 & 0.166 & star-forming dwarf galaxy \\ \hline
        10504-9101 & 0.161 & moderately-inclined disk with inner dust lane \\ \hline
        9889-1902 & 0.159 & barred spiral galaxy with strong star-formation \\ \hline
        8253-12703 & 0.151 & nearly edge-on disk with prominent dust lane \\ \hline
        8145-6102 & 0.148 & several foreground stars \\ \hline
        9880-12703 & 0.145 & possible background galaxy \\ \hline
        8619-12705 & 0.143 & star-forming dwarf galaxy with red, background galaxy \\ \hline
        8988-6102 & 0.141 & moderately-inclined disk with star-forming outskirts and dusty interior \\ \hline
        8440-12705 & 0.133 & edge-on galaxy with prominent dust lane \\ \hline
        8618-3702 & 0.130 & possible foreground star(s) \\ \hline
        9039-9101 & 0.130 & two star-forming galaxies, foreground star \\ \hline
        9029-12701 & 0.127 & edge-on disk with prominent dust lane \\ \hline
    \end{tabular}
    \caption{The twenty galaxies with the largest difference between $M^*$ and $M^*_{\rm LW}$.}
    \label{tab:rogues_gallery}
\end{table*}

%% file: discussion.tex
\section{Discussion and Conclusions}
\label{sec:disc}

In this work, we build upon \citetalias{pace_19a_pca} in this series, which used a library of 40000 spectra of composite stellar populations to construct a reduced-dimensionality space for fitting moderate signal-to-noise optical spectra of nearby galaxies; and as a complement to \citetalias{pace_19a_pca}, investigate the systematics of resolved stellar mass surface density (with respect to kinematically-measured dynamical mass surface density), total (aperture-corrected) galaxy stellar mass (with respect to photometric estimates), and flux-weighted stellar mass-to-light ratios.

We use three \mplv galaxies also observed as part of the DiskMass Survey \citep{diskmass_i} as a testbed for evaluating the systematics of kinematic and spectroscopic estimates of dynamical and stellar mass surface-density. We find that the resolved, inclination-corrected stellar mass surface-density obtained by a simple transformation between PCA-derived $\Upsilon^*_i$ and $M^*$ in most cases exceeds the estimates of dynamical mass surface density. Given that by definition, the dynamical mass includes stars, gas, and dark matter, this discrepancy indicates appreciable systematics in one or both methods. For the two higher-surface-brightness cases, better agreement between dynamical and SPS measurements could be obtained in two cases by postulating that the disk scale-height was overestimated by a factor of $\sim 1.5$. If, though, photometric mass estimates discussed in Section \ref{sec:photometry} were taken as truth (i.e., if the mass estimates here are anomalously heavy by that margin), the scale-height correction could be reduced to approximately a factor of 1.3.

\subsection{Galaxy total stellar-mass: aperture-correction and luminosity-weighting}

In order to estimate total galaxy stellar-mass (including that lying outside the grasp of the IFU), we test two rudimentary methods of aperture-correction: the ``ring" method applies a fiducial mass-to-light ratio equal to the median of the outermost-sampled $0.5 R_e$ to the difference between the NSA K-corrected $i$-band flux and the summed flux from the entire IFU; the favored ``CMLR" method uses the missing flux in the $g$ and $r$ bands to place the remainder of the galaxy on a CMLR. Furthermore, we note that the flux fraction lying outside the IFU is correlated with MaNGA subsample (on average, the Primary+ subsample is exposed out to a smaller galactocentric radius, so the aperture-correction is larger, and in the case of a negative gradient in stellar mass-to-light ratio, the difference between ``ring" and ``CMLR" corrections will be increased). We adopt the ``CMLR" method, since we believe it treats the Primary+ and Secondary samples more equally, and is less susceptible to over-correction.

We also examine the question of whether co-adding an entire IFU's spectra yields a different estimate of average stellar mass-to-light ratio. The intrinsic luminosity-weighting has been attributed to ``outshining" (where young, bright, blue spectra---having intrinsically low mass-to-light ratio---``wash out" the older spectra which sample most of the mass), as well as to the effects of dust. Both scenarios are simply a consequence of luminosity-weighting a mass-related property. We find that for MaNGA IFS data, galaxies with dust lanes or which are viewed edge-on experience the strongest mass-deficits between luminosity-weighted and IFU-summed stellar masses. It is plausible that a different IFS survey with different spatial-sampling characteristics or catalog selection function may experience a different balance of ``outshining" and dust-induced effects. This study is not alone in noting the deleterious effects of spatial binning on the robustness of spectral fits: in tests of the fidelity of full-spectral fitting, \citet{ibarra-medel_avila-reese_19} find that synthetic observations of hydrodynamic simulations of Milky Way-like galaxies with a MaNGA-like instrument, under MaNGA-like observing conditions, and with a spatial binning scheme (intended to increase signal-to-noise ratio) can produce a total mass deficit of up to 0.15 dex. This deficit worsens as inclination increases and a galaxy becomes edge-on: in the extreme, at $i \sim 90^{\circ}$, the light is dominated by the stellar populations at the lowest line-of-sight-integrated optical depth---so, only the outermost ring of stars is seen.

Given the observation that spatial coadding produces biases at the 0.05--0.1 dex level on \logml{i}, it may be prudent to re-evaluate the circumstances under which spectra are automatically coadded. It has become common practice to add together spatially adjacent spectra having low signal-to-noise ratios using adaptive binning techniques like Voronoi binning \citep{cappellari_voronoi}, in order to achieve uncertainties better suited to stellar population analysis. By binning indiscriminate to the properties of the spectra themselves, though, one might group regions in a galaxy that are intrinsically very different. In the case of, for instance, a disk galaxy with star-forming spiral arms and an older bulge or thick disk (having respectively low and moderat-to-high stellar mass-to-light ratios), one spatial bin might include contributions from both a star-forming arm (intrinsically bright) and the nearby bulge or older, thick disk (generally dimmer), potentially biasing an estimate of \logml{i} low. A more reliable approach would preferentially continue to accumulate bins along paths where spectra are similar. One pathway to this might rely on principal component decomposition of observed spectra, followed by agglomeration of nearby spectra that are both nearby and similar in PC space.

\subsection{Public Data}

In addition to the soon-to-be-released resolved maps of \logml{i} described in \citetalias{pace_19a_pca}, we will also release in tabular format estimates of total galaxy stellar-mass as a value-added catalog (VAC) in SDSS Data Release 16 (DR16). Included will be IFU-coadded stellar masses, stellar masses interior to 1 and 2 $R_e$ (where appropriate), and aperture-corrections using the ``ring" and (recommended) ``CMLR" methods described above.

%% file: acknowledgements.tex
\section*{Acknowledgements}
ZJP acknowledges the support of US National Science Foundation East Asia Pacific Summer Institute (EAPSI) Grant OISE-1613857, operated in cooperation with the Ministry of Science \& Technology (MOST) and the China Science \& Technology Exchange Center (CSTEC). ZJP expresses gratitude for the kind hospitality of Nanjing University, where a large fraction of this work took place. ZJP also acknowledges Astro Hack Week 2016, which produced valuable discussions relevant to this work. ZJP and CT acknowledge NSF CAREER Award AST-1554877. Y. C. acknowledges support from the National Key R\&D Program of China (No. 2017YFA0402700), the National Natural Science Foundation of China (NSFC grants 11573013, 11733002). MAB acknowledges NSA Award AST-1517006. This research made use of \texttt{Astropy}, a community-developed core \texttt{python} package for astronomy \citep{astropy}; \texttt{matplotlib} \citep{matplotlib}, an open-source \texttt{python} plotting library; and \texttt{statsmodels} \citep{seabold2010statsmodels}, a \texttt{python} library for econometrics and statistical modeling.

Funding for the Sloan Digital Sky Survey IV has been provided by the Alfred P. Sloan Foundation, the U.S. Department of Energy Office of Science, and the Participating Institutions. SDSS acknowledges support and resources from the Center for High-Performance Computing at the University of Utah. The SDSS web site is www.sdss.org.

SDSS is managed by the Astrophysical Research Consortium for the Participating Institutions of the SDSS Collaboration including the Brazilian Participation Group, the Carnegie Institution for Science, Carnegie Mellon University, the Chilean Participation Group, the French Participation Group, Harvard-Smithsonian Center for Astrophysics, Instituto de Astrof\'{i}sica de Canarias, The Johns Hopkins University, Kavli Institute for the Physics and Mathematics of the Universe (IPMU) / University of Tokyo, the Korean Participation Group, Lawrence Berkeley National Laboratory, Leibniz Institut f\"{u}r Astrophysik Potsdam (AIP), Max-Planck-Institut f\"{u}r Astronomie (MPIA Heidelberg), Max-Planck-Institut f\"{u}r Astrophysik (MPA Garching), Max-Planck-Institut f\"{u}r Extraterrestrische Physik (MPE), National Astronomical Observatories of China, New Mexico State University, New York University, University of Notre Dame, Observat\'{o}rio Nacional / MCTI, The Ohio State University, Pennsylvania State University, Shanghai Astronomical Observatory, United Kingdom Participation Group, Universidad Nacional Aut\'{o}noma de M\'{e}xico, University of Arizona, University of Colorado Boulder, University of Oxford, University of Portsmouth, University of Utah, University of Virginia, University of Washington, University of Wisconsin, Vanderbilt University, and Yale University.